\documentclass{appolb}
\usepackage{epsfig}
\usepackage{graphicx}

\def\bc{\begin{center}}
\def\ec{\end{center}}
\def\beq{\begin{equation}}
\def\eeq{\end{equation}}
\def\bs{\begin{slide}}
\def\es{\end{slide}}
\newcommand{\bmath}{\begin{displaymath}}
\newcommand{\emath}{\end{displaymath}}
\newcommand{\beqn}{\begin{eqnarray}}
\newcommand{\eeqn}{\end{eqnarray}}
\newcommand{\beqns}{\begin{eqnarray*}}
\newcommand{\eeqns}{\end{eqnarray*}}
\newcommand{\ba}{\begin{array}{c}} 
\newcommand{\bat}{\begin{array}{cc}} 
\newcommand{\ea}{\end{array}}

\begin{document}
\title{Chiral Low-Energy Constants~: Status and Prospects
\thanks{IFIC/07-63. Talk given at the 31st International Conference of Theoretical Physics: 
Matter To The Deepest: Recent Development in Physics of Fundamental Interactions,
5-11 Sep 2007, Ustron, Katowice (Poland).}%
}
\author{J. Portol\'es
\address{Instituto de F\'{\i}sica Corpuscular, IFIC, CSIC-Universitat de Val\`encia, \\ Apt. Correus 22085,
E-46071 Val\`encia, Spain}
}
\maketitle
\begin{abstract}
Chiral low-energy constants (LECs) carry the information of short-distance dynamics involving heavier degrees of freedom not present in the chiral Lagrangian. Our knowledge of the LECs is all-important at phenomenological level because their relevance in the prediction of hadronic observables at low-energies and on the other side because they provide hints on the construction of a dual theory of QCD in the low-energy
regime. I review briefly the status of these important couplings.
\end{abstract}
  
\section{Introduction}
Chiral symmetry of massless Quantum Chromodynamics has turned to be a key tool in order to deal with the low-energy domain of strong interactions (typically $E \sim M_{\pi}$) where hadron dynamics is not 
appropriately described by partonic QCD.
It is indeed the guiding principle in the construction of Chiral Perturbation Theory ($\chi PT$)
\cite{Weinberg:1978kz,Gasser:1983yg,Gasser:1984gg} that intends to be the dual effective theory of QCD,
i.e. it describes a perturbative quantum field theory of strong interactions at low-energies.
\par
The $\chi PT$ Lagrangian involves a perturbative expansion guided by powers of external momenta ($p$) and light quark masses ($m$), with $p \sim m$, over a hadronic scale $\Lambda_{\chi}$, driven by the loop expansion required by unitarity, $p^2/\Lambda_{\chi}^2$ or the one that rules the lightest heavier degrees of freedom omitted in the Lagrangian, as the $\rho(770)$ mass, $p^2/M_{\rho}^2$.
It includes the hadronic fields that live in this energy region, the multiplet of Goldstone bosons ($\pi$, $K$, $\eta$), and classical auxiliary fields that help to determine Green functions satisfying the appropriate Ward identities. The theory can be systematically constructed by looking for all the operators  that, with the ingredients above and up to a fixed order in the ${\cal O}(p^n)$ expansion, satisfy the chiral symmetry constraint \cite{Ecker:1994gg}~:
\begin{equation}
{\cal L}_{\chi PT} \, = \, {\cal L}_2^{\chi PT} + {\cal L}_4^{\chi PT} + {\cal L}_6^{\chi PT} + ...  
\end{equation}
${\cal L}_2^{\chi PT}$ embodies the spontaneous breaking of chiral symmetry and depends only on two parameters~: $F \sim 92.4 \, \mbox{MeV}$, related with the decay constant of the pion, and $B_0 F^2 = - \langle 0 | \overline{\psi} \psi |0 \rangle$, the vacuum expectation value of the light quarks \footnote{It happens that the loop expansion of the theory tells us that $\Lambda_{\chi} \sim 4 \pi F$ that is of the same order than $M_{\rho}$. Hence there is only one scale in the perturbative expansion of the theory.}.
\par
Higher orders in the expansion bring in the information of the dynamics of short-distance contributions arising from heavier degrees of freedom that have been integrated out, for instance resonance states. As in any effective field theory (EFT) this information is incorporated into the couplings of the operators~:
\begin{eqnarray} \label{eq:chpt}
{\cal L}_4^{\chi PT} \, = \, \sum_{i=1}^{10} L_i \, {\cal O}_i^{(4)} \, , & \; \; \; \; \; \; \; \; \; \; \; \; 
\; \; \;  \; \; \; & \; 
{\cal L}_6^{\chi PT} \, = \, \sum_{i=1}^{90} C_i \, {\cal O}_i^{(6)} \; , 
\end{eqnarray}
for $SU(3)$. Explicit expressions for the operators can be read from Refs.~\cite{Gasser:1984gg,Bijnens:1999sh}. In Eq.~(\ref{eq:chpt}) $L_i$ and $C_i$ are 
the chiral LECs at ${\cal O}(p^4)$ and ${\cal O}(p^6)$ respectively; they are not provided by chiral symmetry.
At present we do not know how to construct this Lagrangian directly from partonic QCD and, accordingly,
we do not know how to determine the LECs from that framework. Hence we have to use our knowledge on the
foundations of EFTs. LECs in $\chi PT$ should receive contributions from the energy regime at or above the scale that rules the chiral perturbative expansion \cite{Georgi:1991ch}. It is reasonable to infer that the 
dynamics of the lightest meson resonances in the hadronic spectrum, that are not included explicitly in the Chiral Lagrangian, would provide the larger contribution. Following this assumption the determination of the input of the resonance spectrum to the $L_i$ LECs in ${\cal L}_4^{\chi PT}$ 
\cite{Ecker:1988te} showed that they indeed saturate the values extracted from phenomenological analyses. 
As a consequence it is reasonable to think that the most important contribution to the LECs is provided by the energy region immediately above the integrated scale ($E \sim \Lambda_{\chi} \sim M_{\rho}$).
\par
The determination of LECs is crucial for the predictability of low-energy hadronic observables determined
using $\chi PT$. In the last ten years many of these processes have been evaluated up to
${\cal O}(p^6)$ (\cite{Bijnens:2006zp} and references therein) but our ignorance on the values of the involved LECs lessens the practical value of that enormous task. Therefore it is mandatory to explore procedures that allow us to 
determine or at least provide reliable estimates for the values of chiral LECs up to ${\cal O}(p^6)$.

\section{Tools~: Large-$N_C$ and QCD asymptotic constraints}
As illustrated in the ${\cal O}(p^4)$ case \cite{Ecker:1988te,Ecker:1989yg} a procedure to systematically 
disclose the structure of the resonance contributions to the LECs in $\chi PT$ is available. 
The key point is to construct a Lagrangian theory that includes resonances, Goldstone bosons and
auxiliary fields respecting the underlying chiral symmetry. There are several tools that allow us to grab important properties of QCD and to implement them in an EFT-like Lagrangian model. Two relevant features
to consider are~:
\vspace*{0.2cm} \\
i) S.~Weinberg \cite{Weinberg:1978kz} and H.~Leutwyler \cite{Leutwyler:1993iq}
state that if one writes down the most general Lagrangian, including all terms consistent with  
assumed symmetry principles, and then one calculates matrix elements with this Lagrangian to any 
given order of perturbation theory, the result will be the most general possible S--matrix amplitude consistent with analyticity, perturbative unitarity, cluster decomposition and the principles of 
symmetry that have been required. 
\vspace*{0.1cm} \\
ii) The $1/N_C$ coupling ($N_C$ is the number of colours in QCD) can be taken as a perturbative expansion
parameter \cite{'tHooft:1974hx}. Indeed large-$N_C$ QCD shows features that resemble, 
both qualitatively and quantitatively, the $N_C=3$ case \cite{Peris:1998nj}. In practice the consequences of this approach are that meson dynamics in the large-$N_C$ limit
is described by tree diagrams of an effective local Lagrangian involving an infinite spectrum of zero-width mesons.
\vspace*{0.2cm} 
\par
Both statements can be combined by constructing a Lagrangian in terms of $SU(3)$ (Goldstone mesons) 
and $U(3)$ (heavier resonances) flavour multiplets as explicit degrees of freedom respecting the underlying chiral symmetry. Then upon integration of the heavier states, the $\chi PT$ Lagrangian is to be recovered. 
This procedure has been systematically established \cite{Ecker:1988te,Ecker:1989yg,Cirigliano:2006hb} 
and devises what is known as Resonance Chiral Theory ($R \chi T$). Its content is schematically given by~:
\begin{equation} \label{eq:GBdisc}
{\cal L}_{R \chi T} \, = \, {\cal L}_2^{\chi PT} \, + \, \sum_n \, {\cal L}_{n>2}^{GB} \, + {\cal L}_R \; , 
\end{equation}
where ${\cal L}_{n>2}^{GB}$ has the same structure than ${\cal L}_4^{\chi PT}$, ${\cal L}_6^{\chi PT}$, ...
in Eq.~(\ref{eq:chpt}) though with different coupling constants, and ${\cal L}_R$ involves terms with resonances and their couplings to Goldstone modes~:
\vspace*{0.2cm} \\
1) The construction of the operators in the ${\cal L}_R$ is guided by chiral symmetry for the Goldstone
mesons and by unitary symmetry for the resonances. The general structure of these couplings is~:
\begin{equation} \label{eq:structure}
{\cal O} \, = \, \langle \, R_1 R_2 ...R_m \, \chi(p^n)  \, \rangle \, \; \; \; \in \; \; \; \, 
{\cal L}_{(n)}^{\overbrace{\scriptstyle RR...R}^{m}} \; ,
\end{equation}
where  $R_j$ indicates a resonance field and $\chi (p^n)$ is a chiral structured tensor, involving the pseudoscalar mesons and auxiliary fields only. With these settings chiral symmetry is preserved
upon integration of the resonance fields and, at the same time, the low--energy behaviour 
of the amplitudes is guaranteed.  
\vspace*{0.1cm} \\
2) Symmetries do not provide information on the couplings in ${\cal L}_R$ as these incorporate short-distance
dynamics not included explicitly in the Lagrangian. The latter 
is supposed to bridge between the energy region below resonances ($E \ll M_V$) and the parton
regime ($E \gg M_V$). This hypothesis indicates that it should match both regions and it satisfies,
by construction, the chiral constraints. To suit the high-energy behaviour one can match, for instance,
the OPE of Green functions (that are order parameters of chiral symmetry) with the corresponding expressions evaluated within our theory. In addition the asymptotic trend of form factors of QCD currents is estimated  from the spectral structure of two-point functions and it is enforced on the observables. This heuristic strategy is well supported by the
phenomenology \cite{Peris:1998nj,Cirigliano:2006hb,Knecht:2001xc,Cirigliano:2004ue,Cirigliano:2005xn}
and provides information on the resonance couplings.
\par
$R \chi T$ does not have an expansion parameter, hence it does not admit a conventional perturbative
treatment. There is of course the guide provided by $1/N_C$ that translates into the loop expansion, however there is no counting that limits the number of operators with resonances that have to be included in the initial Lagrangian. The number of resonance fields to be kept relies essentially in the physical system that we are interested in. Moreover the maximum order of the chiral tensor $\chi (p^n)$ in Eq.~({\ref{eq:structure}) is constrained by the high--energy behaviour. 
\par
As commented above large-$N_C$ requires, already at $N_C \rightarrow \infty$, an infinite spectrum in order to match the leading QCD logarithms, though we do not know how to implement this in a model-independent way. The usual approach in $R \chi T$ is to include the lightest resonances because their phenomenological relevance, though there is no conceptual problem that prevents the addition of a finite number of multiplets. This cut in the spectrum  may produce inconsistencies in the matching procedure outlined above \cite{Bijnens:2003rc}. To deal with this, one can include more multiplets that may delay the appearance of that problem. 

\section{Status and Prospects}
I comment briefly the present status and several developments on our knowledge of chiral LECs.

\subsection{${\cal O}(p^4)$}
Contributions from the lightest multiplets of vector resonances ($J^{PC}=1^{--}$) to $SU(2)$ \cite{Gasser:1983yg} 
and vector, axial-vector ($1^{++}$), scalar ($0^{++}$) and pseudoscalar ($0^{-+}$) resonances to $SU(3)$ \cite{Ecker:1988te} LECs
are in general good agreement with those values extracted from observables \cite{Amoros:2001cp},
showing that those resonances saturate the phenomenological values. This has been lately
confirmed by the inclusion of one multiplet of resonances with $2^{++}$ and $1^{+-}$ quantum numbers
\cite{Ecker:2007us} that are shown to play, quantitatively, a minor role.
\par
A study of the strange quark mass dependence of the ${\cal O}(p^4)$ $SU(2)$ LECs ($\ell_i^r$) up to ${\cal O}(p^6)$
has also been recently performed \cite{Gasser:2007sg}. The authors essentially obtain~:
\begin{equation}
\ell_i^r \, = \, \delta_{i7} \frac{F^2}{8 B_0 m_s} \, + \, a_i \, + \, b_i \, x \, + {\cal O}(x^2) \; ,
\; \; \; i=1,...,7 \, ,
\end{equation}
where $x=M_K^2/(16 \pi^2 F^2)$ and $m_s$ is the mass of the strange quark. Here $a_i = a_i(L_j,l_K)$
arise at ${\cal O}(p^4)$ and $b_i = b_i(L_j,C_k,l_K)$ at ${\cal O}(p^6)$, with $l_K=\log(M_K^2/\mu^2)$.
The dependence on the strange quark or kaon mass is explicitly stated above. In addition, and for later
discussion, the $b_i$ parameters can be written as $b_i = p_{0,i} + p_{1,i} l_K + p_{2,i} l_K^2$ with
an explicit dependence $p_{0,i}= p_{0,i}(L_j L_k, C_l)$, $p_{1,i} = p_{1,i}(L_j)$.

\subsection{${\cal O}(p^6)$}
In Ref.~\cite{Cirigliano:2006hb} we have constructed the $R \chi T$ Lagrangian needed to evaluate the resonance
contributions to the ${\cal O}(p^6)$ LECs in Eq.~(\ref{eq:chpt}). 
It can be shown \cite{Ecker:1989yg} that, at  ${\cal O}(p^4)$, all local terms in ${\cal L}_4^{GB}$ 
(see Eq.~(\ref{eq:GBdisc})) have to vanish in order not to spoil the asymptotic behaviour of QCD correlators. A corresponding result at ${\cal O}(p^6)$ is still lacking but we have also assumed that all the couplings in ${\cal L}_6^{GB}$ are set to zero.
\par
${\cal L}_{R \chi T}$ in Eq.~(\ref{eq:GBdisc}) involves 124 {\em a priori} unknown couplings. Some
additional work provides an enormous simplification~: \vspace*{0.2cm} \\
i) Upon integration of resonances not all couplings appear independently in the LECs. In general only several 
combination of couplings intervene and to take into account this case one can perform suitable redefinitions of
the fields. This procedure may upset the high-energy behaviour of the theory but it is correct for the evaluation of the LECs. Indeed the unknown couplings are reduced to 77.
 \vspace*{0.1cm} \\
ii) 
The next step is to enforce short-distance information, i.e. the leading behaviour at large momenta, for two and three-point functions and form factors. This procedure, set in Ref.~\cite{Ecker:1989yg}, relies in well-known properties of partonic scattering or asymptotic QCD \cite{Froissart:1961ux}. 
Two-current correlators and associated form-factors provide 19 new constraints on couplings, while the three-point
Green functions studied till now~: $\langle VAP \rangle$ \cite{Cirigliano:2004ue} and $\langle SPP \rangle$
\cite{Cirigliano:2006hb,Cirigliano:2005xn}, give 6 and 5 independent restrictions, respectively. 
We are left  with 47 couplings. Further studies along this line can diminish even more the number
of unknown constants.
\vspace*{0.2cm} 
\par 
Notwithstanding we can already determine fully, at this point, the resonance contribution to the ${\cal O}(p^6)$ couplings $C_{78}$ and $C_{89}$ (that appear in $\pi \rightarrow 
\ell \nu_{\ell} \gamma$ and $\pi \rightarrow \ell \nu_{\ell} \gamma^*$, respectively), $C_{87}$ (in 
$\langle A_{\mu}  A_{\nu} \rangle$), $C_{88}$ and $C_{90}$ (in $F_V^{\pi}(q^2)$ and the $q^2$ dependence of the
form factors in $K_{\ell 3}$), $C_{38}$ (in
$\langle SS \rangle$) and $C_{12}$ and $C_{34}$ (in $F_V^{\pi,K}(q^2)$ and $f_+^{K^0 \pi^-}(0)$). 
\par
The evaluation of resonance contributions to $C_j$ carried out in Ref.~\cite{Cirigliano:2006hb} can be
termed as an $N_C \rightarrow \infty$ evaluation (with a cut spectrum). It is interesting to notice that
when the values obtained for $C_j$ (namely $C_j^R$) are substituted in the expressions obtained in 
Ref.~\cite{Gasser:2007sg} for $p_{0,i}$, the different combinations of $C_j$ couplings vanish (except for $i=7$) \cite{christoph}. Something similar happens with the combinations of $L_i L_j$ products under additional conditions. The reason of these unexpected cancellations is still not understood and it would imply that, in the large-$N_C$ limit, resonance saturation of ${\cal O}(p^4)$ $SU(2)$ chiral LECs is already achieved at leading order. This would be very much satisfactory both for the $1/N_C$ expansion and for the determination of resonance contributions to chiral LECs. It is nice to have something left to understand for the future.

\section*{Acknowledgements}
I wish to thank Micha\l $\,$ Czakon, Henryk Czy\.z and Janusz Gluza for the excellent organization (but for the
weather) of the XXXI International Conference of Theoretical Physics in Ustron (Poland).
This work has been supported in part by the EU MRTN-CT-2006-035482 (FLAVIAnet), 
by MEC (Spain) under grant FPA2004-00996 and by Generalitat Valenciana under grant ACOMP/2007/156.

\end{document}